%% file: main.tex
\documentclass[11pt, a4paper]{article}

\usepackage[top=2cm, bottom=4.5cm, left=2cm, right=2cm]{geometry}
\usepackage{amssymb}
\usepackage{amsmath}
\usepackage{xcolor}
\usepackage{graphicx}
\usepackage{hyperref}
\usepackage{authblk}
\usepackage{tasks}
\usepackage{slashed} 

\hypersetup{%
  colorlinks=true,
  linkcolor=blue,
  linkbordercolor={0 0 1}
}

\newcommand{\GeV}{\text{GeV}}
\newcommand{\TeV}{\text{TeV}}
\newcommand{\invpb}{\text{pb}$^{-1}$}
\newcommand{\pb}{\text{pb}}
\newcommand{\Mb}{\text{Mb}}
\newcommand{\Tb}{\text{Tb}}
\newcommand{\ttbar}{$t\bar{t}$}
\newcommand{\ns}{\text{ns}}
\newcommand{\pt}{p_\mathrm{T}}
\newcommand{\met}{\slashed{E_T}}

\begin{document}

\input{body}

\end{document}

%% file: body.tex
\begingroup\centering
{\Large\bfseries\mathversion{bold} The terascale tutorial}%

\vspace{7mm}
 
 \begingroup\scshape\large
 Konstantinos Theofilatos\footnote{comments to konstantinos.theofilatos@cern.ch, http://www.cern.ch/theofil}\\
 \endgroup
 \vspace{2mm}
 \begingroup\small
 Department of Physics\\
 National and Kapodistrian University of Athens\\ 
 \endgroup
 \vspace{10mm}

\textbf{Abstract}\vspace{5mm}\par
\begin{minipage}{14.7cm}
This note summarizes the lectures given in the tutorial session of the \emph{Introduction to the Terascale} school at DESY on March 2023. 
The target audience are advanced bachelor and master physics students. 
The tutorial aims to best prepare the students for starting an LHC experimental physics thesis.
The cross section of \ttbar\ pair production is detailed alongside with the reconstruction of the invariant masses of the 
top quark as well as of the $W$ and $Z$ bosons. 
The tutorial uses ideas and CMS open data files from the \texttt{CMS HEP Tutorial} written by C. Sander and A. Schmidt, 
but is entirely rewritten so that it can be run in {\texttt Google Colab Cloud} in a columnar style of analysis with python. 
In addition, a minimal {\texttt C/C++} version of a simple event-loop analysis relying on \texttt{ROOT} is exampled. 
The code is kept as short as possible with emphasis on the transparency of the analysis steps, rather than the elegance of the software, 
having in mind that the students will in any case need to rewrite their own custom analysis framework. 
\end{minipage}\par
\endgroup

\newpage
\tableofcontents
\newpage

\section{Introduction}
In an learning by example approach, we will discuss how to measure the cross section of a physics process, which is known as top quark anti-quark pair $(t\bar{t})$  production.

\begin{figure}[ht]
\centering
\includegraphics[width=0.5\textwidth]{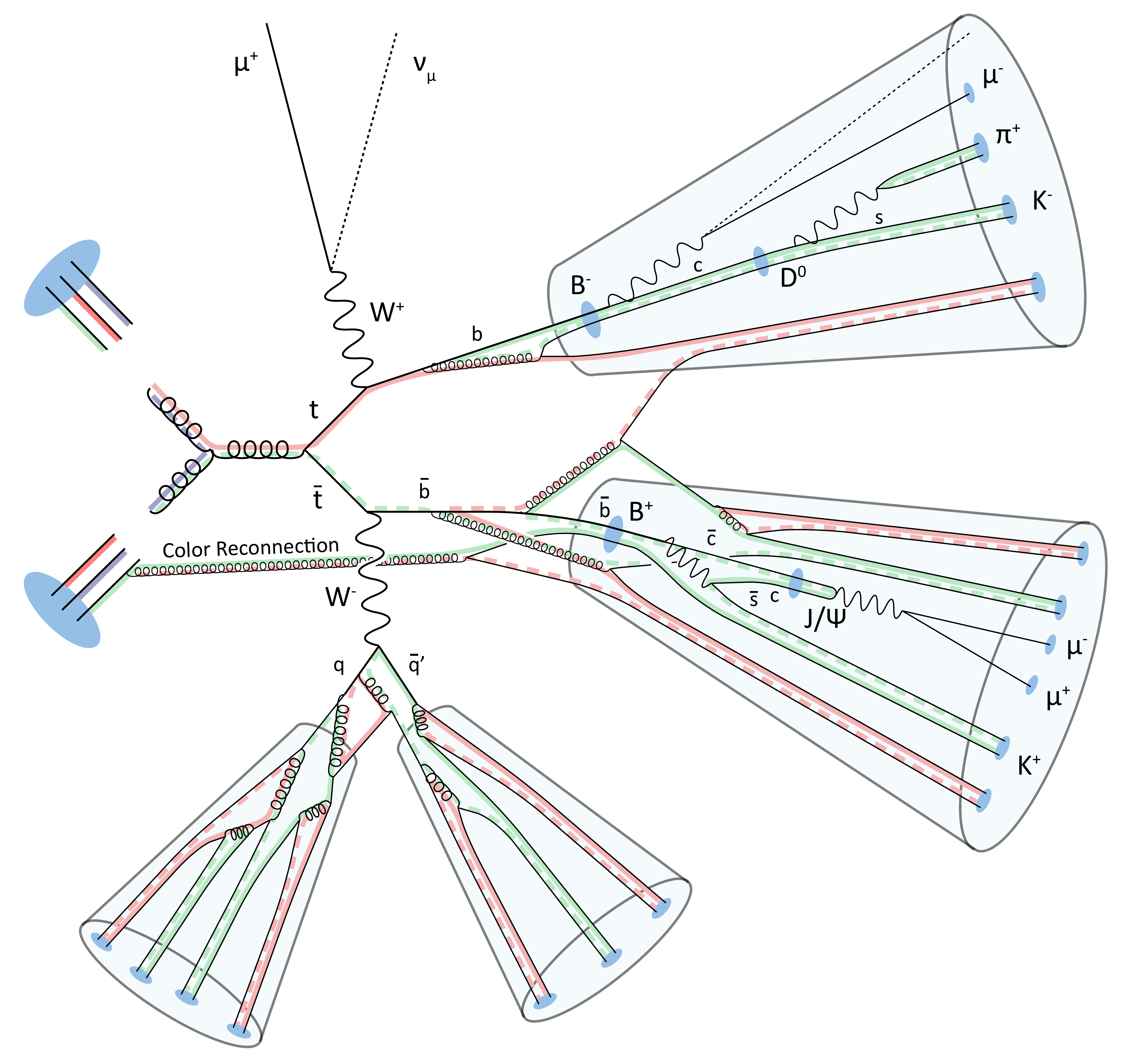}
\caption{Artistic visualization of a \ttbar\ produced by two colliding protons (on the left in blue) 
and decaying into a $\mu^{-}$ and hadrons that are later on clustered as jets. (Image credits: B. Stieger.)}
\end{figure}

We will make use of a pocket-size data sample that comes with the {CMS HEP Tutorial}~\cite{heptutorial}, comprising of just a small fraction of $pp$ collision data of 50~\invpb at $\sqrt{s} = 7~\TeV$.
All data \& MC simulation files as well as the code can be found in:

\begin{itemize}
\item \href{http://theofil.web.cern.ch/theofil/cmsod/files/}{http://theofil.web.cern.ch/theofil/cmsod/files/}
\item \href{https://github.com/theofil/I2TheTerascale}{https://github.com/theofil/I2TheTerascale}
\end{itemize}

The data have been selected among the many $pp$ collisions occurring every second at the LHC, such as at least one muon is present in the collision debris. 
This type of selection has been made using the so-called single muon trigger of the CMS detector (see Sec.~\ref{sec:trigger}). 
Instead of a lengthy intro on the LHC and how a particle physics detector works, few video links below that need to be appreciated before moving forward.
\begin{itemize}
\item \href{https://www.youtube.com/watch?v=pQhbhpU9Wrg}{LHC YouTube video, absolute must see!}
\item \href{https://www.youtube.com/watch?v=S99d9BQmGB0}{CMS YouTube video, all what you need to know for getting started}
\end{itemize}
In addition, a very nice introduction for collider physics has been written by M. D. Schwartz~\cite{tasi}, should the students wish to dive deeper into the physics.

\section{Physics Analysis}
The most basic quantity we are interested in particle physics is called \href{https://en.wikipedia.org/wiki/Cross_section_(physics)}{cross section} $(\sigma)$ for a particular particle 
interaction to occur. 
You could think the cross section of a process as the analogous of the probability for that process to take place, but instead being a pure number it is measured in units of area, 
\href{https://en.wikipedia.org/wiki/Barn_(unit)}{1 barn} $= 10^{-28}m^{2}$.

The sample size of the LHC $pp$ collision data is quantified by what is known as (integrated) luminosity measurement $(L)$ and has units of inverse area (e.g., \invpb), 
where $p$ stands for the pico $=10^{-12}$ order of magnitude. Smaller cross section area implies smaller chance for the interaction to occur. 
On the other hand, more $pp$ collisions on tape, means more L. 
So we can probe a process of small $\sigma$ if $L$ is sufficiently large, provided that we have a way to select $pp$ events enriched with the process of interest  
as in most cases a $pp$ collision results into a ``boring'' final state.  

\subsection{The master equation: \texorpdfstring{$N = \epsilon \sigma L$}{N = epsilon sigma L}}
The number of events $(N)$ we expect for a specific process  with known cross section $(\sigma)$ in a data sample of known (integrated) luminosity $(L)$ is:
\begin{equation}
N = \epsilon \sigma L
\label{eq:master}
\end{equation}
where $\epsilon$ is the total selection efficiency for recording this process, including both kinematic and geometric acceptance of the detector.

\subsection{Physics Processes}
While we have some control of the initial state, e.g., the center of mass energy of the colliding protons, we don't really control what comes out in the final state. It is like rolling a dice with an unknown number of faces and different frequencies for each of the possible outcomes.
Provided that there is sufficient energy in the initial state, all possible paths (particle interactions) will be taken 
by nature with probabilities that governed (we believe) by the laws of quantum mechanics. 
During LHC Run II,  for $\sqrt{s} = 13$~\TeV\ and $20$~nb$^{-1}/s$ {\em instantaneous} luminosity the production rate for different physics processes is shown below. 

\begin{table}[h]
\begin{center}
\begin{tabular}{ rl } 
 process          & rate (Hz) \\\hline
 $W^{\pm}$        & 4000      \\\hline    
 $Z^{0}$          & 1200      \\\hline 
 $t\bar{t}$       & 17        \\\hline
 $h^{0}$          & 1         \\\hline
 $h^{0}h^{0}$     & (0.007 ?)  \\\hline  
\end{tabular}
\end{center}
\caption{Expected production rate of different processes at the LHC Run II with $\sqrt{s} = 13$~\TeV\ and $20$~nb$^{-1}/s$ instantaneous luminosity. 
 The last process has yet been confirmed and is one of the main goals of the LHC as it is particularly sensitive to the Higgs self-coupling.
}
\label{tab:rates}
\end{table}

In fact, those particles are produced by nature in an effortless manner for the given instantaneous luminosity and center-of-mass energy of the $pp$ collisions. 
However, the particle detectors don't detect directly the very short lived particles listed above, but rather their decay products.
We simply cannot speak at an event-by-event level that this event is Higgs, this is a $W$ and so on, although surely we will hear people saying so when they look into beautiful event displays. 

By applying selection criteria (analysis cuts) on the $pp$ data, one can increase the efficiency of selecting a specific process (call it signal: $S$) 
against other processes (call them backgrounds: $B$) that will also satisfy the applied criteria mimicking the signal.
Ideally, we would want the signal efficiency to be $100\%$ while the backgrounds to have $0\%$ efficiency.
Unfortunately, this is almost never the case and there is always some background contribution in the sub-sample of data we selected to focus our attention. The amount of background events in our signal-enriched sample has to be estimated and MC simulation might be used for that purpose.
It is therefore typical that together with the MC simulation of the signal we do also consider the background simulation, which is usually much more difficult to get correctly 
(i.e., having larger uncertainty on the predicted event yields due to theory uncertainties in its $\sigma$). 

\section{Event Weights, MC and Statistical Uncertainty}
In practice, multiple processes contribute to the signal region, the number of events expected from MC $(N_\mathrm{MC})$ is
\begin{equation}
N_\mathrm{MC} = \sum_\mathrm{i} \epsilon_\mathrm{i} \sigma_\mathrm{i} L
\end{equation}
where the index $i$ enumerates all simulated physics processes. 
Without any event selection, the total number of events that have been generated for the processes $i$ is
\begin{equation}
N_\mathrm{MC,i}^\mathrm{tot} = \sigma_\mathrm{i} L_\mathrm{i}
\end{equation}
where $L_\mathrm{i}$ is the simulated luminosity for the specific sample, which in general varies as function of the total number of computing hours used for the MC generation.
In order to normalize all samples to the luminosity of data $L = 50$~\invpb, we need to assign them appropriate weights
\begin{equation}
w_\mathrm{i} = \frac{\sigma_\mathrm{i} L} {N_\mathrm{MC,i}^\mathrm{tot}} = \frac{L}{L_\mathrm{i}} 
\end{equation}
where the index $i$ is, as before, enumerating the simulated physics processes.
To give an example, if $w_\mathrm{i} = 5$ we would need to count each entry (1 unweighted event) of the MC sample as 5 weighted MC events, when comparing simulation with data.
On the contrary, if $w_\mathrm{i} = 0.1$ we need to count every 10 entries (10 unweighted events) of the MC as 1 weighted MC event.
The statistical uncertainty of weighted (Poisson in nature) MC events is not just as simple $\sqrt{N}$ but is rather given by 
\begin{equation}
\label{eq:sumw2}
\delta N_\mathrm{MC}^\mathrm{sel} = \sqrt{\sum_{j} w_{j}^2}  
\end{equation}
where now the index ${j}$ counts all entries (unweighted events) of the MC processes $(\mathrm{i})$ that contribute to a desired event selection.

\subsection{Exercises}
Assuming that we have B = 1000 (weighted) MC events, when applying the event selection of our signal region.
Calculate what would be $\delta B/B$ if
\begin{enumerate}
\item all MC events have $w_{j} = 0.1$
\item all MC events have $w_{j} = 10$
\end{enumerate}
assuming that $\delta B$ is dominated by uncertainties of statistical nature, neglecting systematic uncertainties.

\section{Data and MC samples}

\begin{table}
\begin{center}
\begin{tabular}{ rlr } 
 process   & $\sigma [\pb]$ & triggerBit    \\\hline\hline
 data      & --         & always true   \\\hline 
 TTbar     & 165        & true or false \\\hline 
 WJets     & 31300      & always true  \\\hline 
 DYJets    & 15800      & always true\\\hline 
 WW        & 4580       & always true\\\hline 
 WZ        & 3367       & always true\\\hline
 ZZ        & 2421       & always true\\\hline
 SingleTop & 5684       & always true\\\hline
 QCD       & $\sim10^8$ &always true \\\hline
\end{tabular}
\end{center}
\caption{Cross section and trigger information for the MC samples~\cite{heptutorial}.}
\end{table}

In total for this tutorial, we have $N = 469384$ data events satisfying the single muon trigger, for an integrated luminosity of $50$~\invpb of $pp$ collisions at $\sqrt{s} = 7$~\TeV.
Ideally, we would have wanted to have at least $\times 10$ MC simulated events (i.e., $500$~\invpb of simulated luminosity), 
in order for the MC statistical uncertainty to be less than the one of the data. 
If that would have been the case, assign each MC event a weight of $w = 0.1$, i.e., counting each entry found in MC as $0.1$ events. 
Unfortunately, this is not possible for processes with very large cross section where in practice we are only able to simulate much less events than those expected for $50$~\invpb.
For these processes, the simulated luminosity is smaller than $50$~\invpb. 
Below follow the available weighted and unweighted events for data and all MC processes we will use in analysis.

\begin{verbatim}
Data: 469384.0 ± 685.1   [entries: 469384]
MC  : 331407.3 ± 55461.7 [entries: 240601]
-----------------------------
WJets  209576.7 ± 689.2  [entries: 109737]
DYJets  34113.2 ± 145.6  [entries: 77729]
TTbar  7928.6 ± 45.5     [entries: 36941]
WW  229.9 ± 3.7          [entries: 4580]
WZ  69.9 ± 1.3           [entries: 3367]
ZZ  16.9 ± 0.4           [entries: 2421]
Single Top  311.6 ± 4.4  [entries: 5684]
QCD  79160.5 ± 55457.2   [entries: 142]
-----------------------------
\end{verbatim}

The first number corresponds to the number of weighted events and their uncertainty, while in $[entries: ...]$ the number of entries (or unweighted events if you wish) is given.
By construction we have $w = 1$ for data and the number of weighted events is equal to the number of unweighted events (entries) in this case.
Note that when $w=1$, the statistical uncertainty given by Eq.~\ref{eq:sumw2} reduces to $\sqrt{N}$.
The QCD background has by far the largest event weight, for which $142$ entries (unweighted events) correspond to as much as $79k$ events with very large statistical uncertainty $\sim 55k$.
Already at this point, we get warned that this type of background should be filtered away by some event selection (cuts). 

\subsection{Exercises}
If we define as our signal the final state of $t\bar{t} \to b \bar{b} q \bar{q} \mu^{+} \nu$, sketch in a piece of paper possible ways for which the background simulated process (DY+jets, TTbar, WW, WZ, ZZ, SingleTop, QCD) 
can mimic our signal.

\section{Opening ROOT files}
The \href{https://www.home.cern/science/accelerators/large-hadron-collider}{LHC} experiments use \href{https://root.cern}{ROOT} to analyze and {\em store} the information recorded during 
hadronic collisions. So, if we would want to study the interactions taking place during $pp$ collisions, we need to learn how to read the ROOT files produced by the experiments.
There are many ways to open a ROOT file, the most popular are:

\begin{enumerate}
\item Install \href{https://root.cern}{ROOT}\footnote{For Windows, see also \href{https://github.com/theofil/I2TheTerascale/raw/main/docs/ROOT_Windows_InstallationFromSource.pdf}{these instructions} in case you do not find your way with the official ones.}
\item Install \href{https://uproot.readthedocs.io/en/latest/}{uproot} and \href{https://awkward-array.readthedocs.io/en/stable/index.html}{awkward} arrays.
\end{enumerate}

In addition a third way we developed here, is to use the {\em Google Colab} suite and install there all python packages needed to run an analysis.  
This approach is the least optimal but has the fastest time-to-analysis for the students, since it does not need any installation to a local computer.
See how this works, by opening \href{https://github.com/theofil/I2TheTerascale/blob/main/code/python/openROOTFile.ipynb}{openROOTFile.ipynb}. 
The code above can be easily modified to open and analyze any ROOT file.

\subsection{Exercises}
\begin{enumerate}

\item Open ``data.root'' from \href{http://theofil.web.cern.ch/theofil/cmsod/files/}{http://theofil.web.cern.ch/theofil/cmsod/files/}
\item Check what's inside the ``TTree'' named ``events''.
\item Count how many events have exactly 1 $\mu$ and at least 2 jets.
\end{enumerate}

\section{What's inside the ROOT files?}
\begin{figure}[ht]
\label{fig:cms}
\centering
\includegraphics[width=0.7\textwidth]{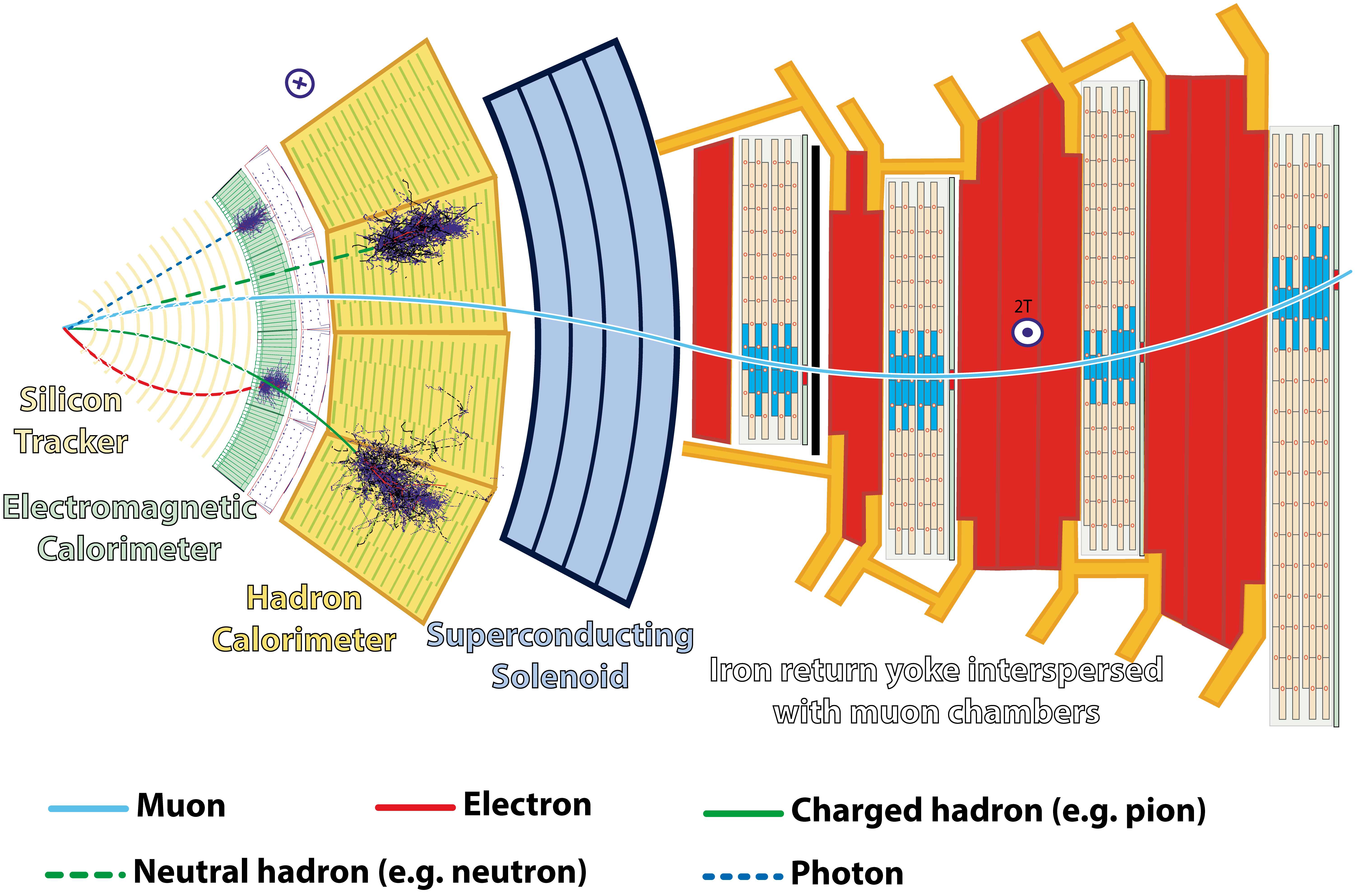}
\caption{Particles seen by the CMS detector.}
\end{figure}

Inside the ROOT file we can find all the information needed to build Lorentz four-vectors 
\begin{equation*}
p^{\mu} = (E, p_\mathrm{x}, p_\mathrm{y}, p_\mathrm{z})
\end{equation*} 
of the particles detected by the CMS detector (Fig~\ref{fig:cms}).
We assume as known the masses, of muons, electrons as well of the pions. We measure their momenta $(\vec{p})$ using the deposits they leave as they go through the detector. 
Particles are also grouped into jets 
\begin{equation*}
p^{\mu}_\mathrm{jet} = \sum_\mathrm{i}  p^{\mu}_i
\end{equation*} 
using a clustering algorithm to decide which particles ($i$) will be grouped together.
The particle jets are usually interpreted as the evolution of the partons $(q,g)$ produced in the hard scatter, but it should be kept in mind that 
their four-momenta is not $1-1$ even in MC truth, due to the QCD color confinement as well as the ambiguities arising from the clustering itself.

The transverse momentum imbalance, with its magnitude best known as missing transverse energy $\met$ or simply MET, is defined as  
\begin{equation*}
\vec{p}_\mathrm{Tmiss} =  - \sum_\mathrm{i} \vec{p}_{\mathrm{T,i}}
\end{equation*} 
where the index $i$ (usually) runs over all visible particles that satisfy the experimental thresholds and pass some predefined identification criteria.
More information on the contents of the ROOT files can be found here~\cite{heptutorial}

While the naming convention might be different, other data formats storing information by ATLAS and CMS typically give access to similar type of information. 
Getting familiar on how to use the ones given here, 
makes evident how to do the same type of job with other types of data.

\subsection{Exercises}
Go to the \href{https://atlas.cern/Resources/Opendata}{ATLAS open data} and \href{https://opendata.cern.ch/docs/about-cms}{CMS open data}, find
your favorite dataset and open it using a modified version of the \href{https://github.com/theofil/I2TheTerascale/blob/main/code/python/openROOTFile.ipynb}{openROOTFile.ipynb}.

\section{Data vs MC, histograms and histogram stacks}
The most standard way to compare Data/MC is to make histograms for observables of interest.
An example variable of interest here, is the muon multiplicity
\begin{figure}[ht]
\centering
\includegraphics[width=0.5\textwidth]{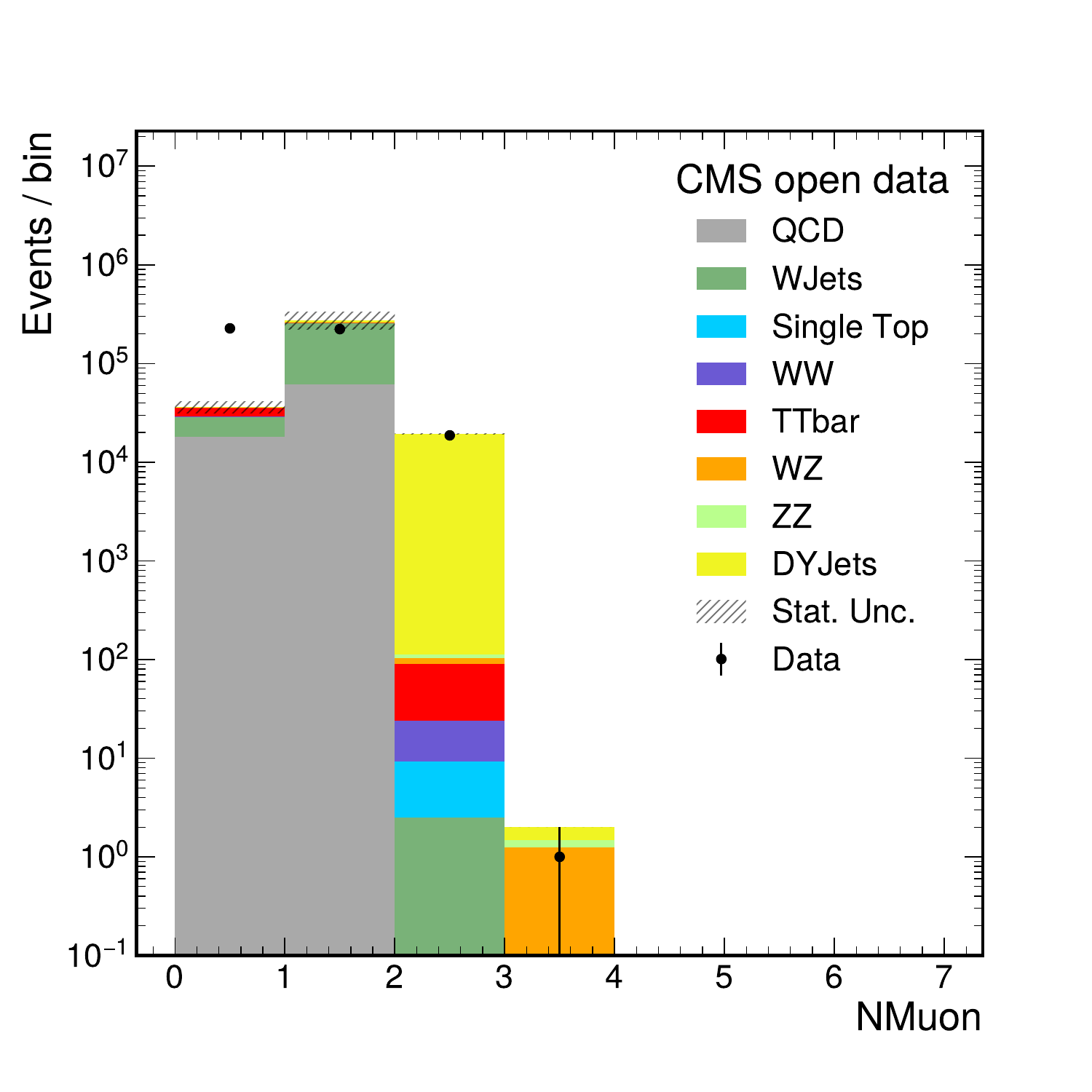}
\caption{Muon multiplicity from the data and MC files used in the \texttt{CMS HEP Tutorial}.}
\label{fig:nmuon}
\end{figure}
in our data and how they compare with the MC simulation (Fig.~\ref{fig:nmuon}).
The events data are binned in a histogram counting how many (offline muons) are present in each of our events $N=469384$. 
Conventionally the data histogram is shown with black circular points (or sometimes squares) with $\sqrt{N}$ error bar if they are (pure) event counts.
Events from each of the  MC are binned in separate histograms with different colors and then stacked on top of each other to compute the expected event yield from simulation.
All MC processes are normalized to the luminosity of data and each event has its own event weight. The statistical uncertainty of the MC estimation is then estimated by Eq.~\ref{eq:sumw2}.

We can experiment on making such graphics using:
\begin{itemize}
\item \href{https://github.com/theofil/I2TheTerascale/blob/main/code/C/makePlot.C}{makePlot.C}
\item \href{https://github.com/theofil/I2TheTerascale/blob/main/code/python/LazyHEPTutorialColab.ipynb}{LazyHEPTutorialColab.ipynb} 
\end{itemize}

Doing such graphics synopsizes all the event counts we have in Data and MC, in a very economic manner. 
But we should not forget that our program knows more details and we should be able to be more verbose if required.
We do this once for Fig.~\ref{fig:nmuon}, as an example.

\begin{verbatim}
### printing number of events for each bin and its estimated uncertainty ###
###       disable this if you wish by setting printOut = False           ###

Data [ 227265.0,  223411.0,  18707.0,  1.0,  0.0,  0.0,  0.0,  ]
DataError [ 476.7,  472.7,  136.8,  1.0,  0.0,  0.0,  0.0,  ]
MCTot = [ 36534.6,  275505.5,  19365.2,  2.0,  0.1,  0.0,  0.0,  ]
MCTotError = [ 5041.7  55231.1  103.7  0.0  0.0  0.0  0.0  ]

### detailed analysis of MC ###

QCD = [ 18058.3,  61102.2,  0.0,  0.0,  0.0,  0.0,  0.0,  ]
QCDError = [ 5039.1,  55227.8,  0.0,  0.0,  0.0,  0.0,  0.0,  ]
WJets = [ 11070.9,  198503.2,  2.5,  0.0,  0.0,  0.0,  0.0,  ]
WJetsError = [ 154.8,  671.6,  2.2,  0.0,  0.0,  0.0,  0.0,  ]
Single Top = [ 13.8,  291.1,  6.7,  0.0,  0.0,  0.0,  0.0,  ]
Single TopError = [ 0.9,  4.3,  0.6,  0.0,  0.0,  0.0,  0.0,  ]
WW = [ 11.6,  203.8,  14.6,  0.0,  0.0,  0.0,  0.0,  ]
WWError = [ 0.8,  3.5,  0.9,  0.0,  0.0,  0.0,  0.0,  ]
TTbar = [ 6589.8,  1272.8,  65.9,  0.0,  0.0,  0.0,  0.0,  ]
TTbarError = [ 41.4,  18.3,  4.2,  0.0,  0.0,  0.0,  0.0,  ]
WZ = [ 2.6,  52.2,  13.9,  1.2,  0.0,  0.0,  0.0,  ]
WZError = [ 0.3,  1.1,  0.6,  0.2,  0.0,  0.0,  0.0,  ]
ZZ = [ 0.3,  6.5,  9.8,  0.2,  0.1,  0.0,  0.0,  ]
ZZError = [ 0.0,  0.2,  0.3,  0.0,  0.0,  0.0,  0.0,  ]
DYJets = [ 787.3,  14073.6,  19251.8,  0.5,  0.0,  0.0,  0.0,  ]
DYJetsError = [ 23.5,  99.9,  103.2,  0.5,  0.0,  0.0,  0.0,  ]
\end{verbatim}

Already at this point we see that the QCD MC sample is a trouble maker. It contributes many events, with huge relative uncertainty. Even worse, MC is in large disagreement with data for the first two bins. We would like to restrict our analysis in a suitable subsample, applying an event selection that will eliminate the QCD contribution hoping that the Data/MC will become more reasonable.

We can experiment with the code to make it select only events for which the single muon trigger has  fired\footnote{\texttt{triggerIsoMu24 == true, see Sec.~\ref{sec:trigger}}} and 
the muon (offline) transverse momentum is above  $24$~\GeV, which is the trigger threshold, requiring that $\pt>25$~\GeV. 
In addition, we can further restrict the event selection, requiring that in parallel jets and other physics objects are present in the final state and compare again Data/MC for that subsample.
 
\subsection{Exercises}
\begin{itemize}
\item[a)] Study the muon $\pt$ distribution in bins of $1$~\GeV\ for the range $0<\pt<50$~\GeV.
\item[b)] Repeat a) for when \texttt{triggerIsoMu24} is true
\item[c)] Remake the muon multiplicity for when the muon $\pt>25$~\GeV\ and trigger is fired.
\end{itemize}

\section{Triggering}
\label{sec:trigger}
The LHC is designed to produced almost 1 $pp$ bunch crossing event every $25$~\ns. In RAW format, the event size is $O(1)$~\Mb. 
Recording all the $pp$ collision events would require to write on tape $\sim40$~\Tb/$\mathrm{s}$.
It is thus a necessity for ATLAS and CMS to use an almost real time (online) decision system for selecting which of the LHC bunch crossing are the most interesting ones to be kept on tape for
offline analysis. 

The first level of selection is known as Level-1-trigger and is made by very fast algorithms encoded in FPGAs. 
The L1 algorithms have to be quicker than $3.2\mu s$ and do partial and coarse reconstruction of physics objects like (jets, e/$\gamma$, MET, $\mu$, $\tau$, $b$-jets ...) that are used to decide 
if the event will be kept on tape for offline analysis. Events that are not firing the L1-trigger, are lost for ever. 
Further qualification criteria are imposed by algorithms running on a computing farm, known as High Level Trigger. Events that have passed the two levels of triggering 
(L1 and HLT) are available for offline studies, where typically the person doing the analysis defines further (offline) criteria to define how the signal should look like.

A data event has to be ``triggered'' to be kept on tape. The selection efficiency of the triggering system for the signals of interest, is of great importance 
and is among the dominant experimental uncertainties.
In MC simulation, we can emulate the trigger system and study the triggering efficiency for the physics signals we are interested into. 
For that we will need MC simulated data (mock data) that include also events that in reality would not pass the trigger requirements. 
Here, the corresponding event flag accompanying each event entry is named ``triggerIsoMu24`` and is available as a TRUE (1) or FALSE (0) bit inside the data and MC root files. 
However only for the \ttbar\ MC events with ``triggerIsoMu24==0'' are available\footnote{By construction real data have always ``triggerIsoMu24==1'' here, while the rest of the MC simulated processes
but the \ttbar\ have been filtered using the ``triggerIsoMu24==1'' to reduce the size of the data sample.}.

\subsection{Exercises}
\begin{enumerate}
\item Open ``ttbar.root'' from \href{http://theofil.web.cern.ch/theofil/cmsod/files/}{http://theofil.web.cern.ch/theofil/cmsod/files/} and measure the efficiency of the ``triggerIsoMu24==1'' selection. 
\item Repeat the efficiency measurement as function of the generated $\pt$ of the $\mu$.
\item Quantify the MC statistical uncertainty of the measured efficiencies.
\end{enumerate}


\section{Cross Section Measurements}

Perhaps the most fundamental type of an LHC physics analysis is a cross section measurement. This can be found turning around Eq.~\ref{eq:master}.
We measure $L$ from data and estimate $\epsilon$ for the signal $(S)$ using MC simulation (sometimes corrected with data-driven scale factors).
  
Assuming that MC predicts with good accuracy the amount of background we expect in the signal region ($B$), the measured signal yield in data should be just $N - B$. In the ideal case (with no uncertainties of any type) we would expect by construction that $N = S + B$, with $N$ being the measured event counts in the signal region of data and $S$ and $B$ the expected signal and background in the signal region, which here we will get solely from MC simulation.
Dividing our signal candidate events in data $(N-B)$ by the factor $\epsilon L$, gives an estimate of signal's cross section in data, which could be compared with the $\sigma$ expected from theory.
In its simplest incarnation an LHC cross section measurement\footnote{A more sophisticated approach would be to minimize the likelihood (fit) of all signal and background samples constrained taking into account normalization and shape uncertainties.}
is as simple as an event counting experiment, provided that we know accurately $B$, $\epsilon$ and $L$.

The uncertainty of $L$ is at the level of $2-3$ percent, so all the analysis challenge boils to finding a way to define the signal region
such that the uncertainties $\delta\epsilon$ and $\delta B$ come out small, while (ideally) $\epsilon\to 1$ and $B\to 0$. 
Nowadays, neural networks of several hidden layers are used to define the event selection of the signal region, 
but keep in mind that the increase in sensitivity might come with increased systematic uncertainty, which is nontrivial to estimate.

\subsection{Exercises}
Measure the $t\bar{t}$ cross section for a signal region that you will define, to achieve good signal significance using the  $S/\sqrt{B}$ as figure of merit. 
To calculate the signal selection efficiency $\epsilon$ we need to count how many $t\bar{t}$ (weighted) events we have at our disposal in total 
($N_\mathrm{gen}^{\mathrm{tot}}$) inside the \texttt{ttbar.root} file. 
The efficiency will simply be $\epsilon = N_\mathrm{sel}/N_\mathrm{gen}^{\mathrm{tot}}$, where $N_\mathrm{sel}$ is the total number of (weighted) events 
passing the selection cuts of our signal region.

Assume that the relative uncertainty  for the signal selection efficiency is $30\%$  (i.e., $\delta\epsilon/\epsilon = 0.3$) and that the luminosity $L$ comes with $5\%$ uncertainty. 
For the background estimation $B$, assume that is only as large as the corresponding MC statistical uncertainty reported by your program.
Compare your measurement with the first measurement that CMS ever made~\cite{CMSXS}, using pretty much the same data. 
What's the main differences among them and how they compare with yours in terms of precision?

\section{Projects}

\subsection{Trigger efficiency as function of the \texorpdfstring{$\mu$ $\pt$}{mu pt}}
Study the trigger efficiency of the signal, defined here as the semileptonic decay of $t\bar{t}$ pairs, leading to final states with $N_\mathrm{\mu}>=1$. 
For this project you will mostly need to open just the \texttt{ttbar.root} , as it's the only sample we have available for which the events that do not 
pass the \texttt{triggerIsoMu24} bit, i.e.,  events having \texttt{triggerIsoMu24 == false} are also stored inside the \texttt{ROOT} file. 

Calculate $\epsilon_\mathrm{trigger} = \mathrm{pass}/\mathrm{total}$  in bins of the generated muon (MC truth) $p_\mathrm{T}$. 
Select events that have one muon generated \texttt{fabs(MCleptonPDGid) == 13} and calculate the MC generated $p_\mathrm{T}$ of the muon using the \texttt{MClepton\_px} and 
\texttt{MClepton\_py} branches. Estimate the efficiency for a generated muon to pass the CMS trigger ``triggerIsoMu24 == true`` as a function of its $p_\mathrm{T}$  (i.e., in bins of pt),  starting with very fine binning e.g., 0.25 or 0.5 GeV in width and increasing it to 1-20 GeV widths at high $p_\mathrm{T}$ for when the available statistics start to be an issue.

Thinking needs to be placed for what would be the statistical uncertainty of $\epsilon_\mathrm{trigger}$. For simplicity we can calculate the uncertainty on the efficiency estimation, in the normal frequentist approximation.  In this approximation we assume that  the observed events that pass the selection $(n)$ over the total events $(N)$, $\epsilon = n/N$, is an estimate of the true efficiency 
$\epsilon_\mathrm{true}$. 
They uncertainty in the estimated efficiency in the normal approximation and the large-N limit is $\delta\epsilon = \sqrt{(\epsilon(1-\epsilon)/N)}$~\cite{binomialCI}. 
\begin{enumerate}
\item
Explain why this definition of $\delta\epsilon$ is reasonable, starting from the fact that $n\sim \mathrm{binomial} (N, p = \epsilon_\mathrm{true})$ and that we approximate the unknown true efficiency 
$p = \epsilon_\mathrm{true} \approx \epsilon = n/N$.
 
\item Furthermore,  verify that the branching ratio we get in ttbar MC for events with exactly $1\,\mu$ and no other charged leptons in the final state (semi-leptonic final state in the muon channel), is what we expect given that $BR(W\to\mu\nu) \approx 10.6\%$.
\end{enumerate}

Key figures to study:
\begin{itemize}
\item $\epsilon_\mathrm{trigger}$ in bins of the generated muon $p_\mathrm{T}$, when taking into account the event weights but ignoring any uncertainty.
\item  $\epsilon_\mathrm{trigger}$ in bins of the generated muon $p_\mathrm{T}$, without taking into account the event weights and estimating the corresponding uncertainty in the normal frequentist method.
\item reconstructed muon $p_\mathrm{T}$ calculated from \texttt{Muon\_Px[0]} and \texttt{Muon\_Py[0]} for data and MC, without any threshold in the muon $p_\mathrm{T}$ for events selected 
\texttt{triggerIsoMu24 == true}. (For this plot you will need to modify \texttt{makePlot.C} analysis script and use fine binning of 1 \GeV\ width.)
\end{itemize}

\subsection{\texorpdfstring{$W$}{W} control region and the \texorpdfstring{$W$}{W}-boson transverse mass}
The event preselection starts with requiring exactly one muon $(N_\mathrm{\mu} = 1)$ final state, for events with \texttt{triggerIsoMu24==1} true. 

Study the reconstructed muon $p_\mathrm{T}$ calculated from \texttt{Muon\_Px[0]} and \texttt{Muon\_Py[0]}  without any threshold starting from $p_\mathrm{T}=0$, 
for events selected \texttt{triggerIsoMu24 == true}. (For this plot you will need to modify \texttt{makePlot.C} analysis script and use fine binning of 1 GeV width.)  Show that is reasonable to select only those events with a leading muon having  $p_\mathrm{T}>25$ GeV. 

For the selected events (i.e., preselection + $p_\mathrm{T}>25$ GeV requirement), produce the transverse mass $m_\mathrm{T}$ and the MET  distributions as well as the $(N_{j})$ and b-jet $(N_\mathrm{bj})$ multiplicity distributions. Key figures:
\begin{itemize}
 \item reconstructed muon $p_\mathrm{T}$ for data and MC, without any threshold in the muon $p_\mathrm{T}$ for events with $N_\mathrm{\mu}==1$ that fire the trigger, in bins of 1 GeV width.
 \item MET in bins of 10 GeV width
 \item transverse mass $m_\mathrm{T}$ in bins of 10 GeV width
 \item jet multiplicity $N_{j}$
 \item b-jet multiplicity $N_\mathrm{bj}$
 \item event counting statistics summary
\end{itemize}

\subsection{Drell--Yan control region and the \texorpdfstring{$Z$}{Ζ} boson mass}
The event preselection starts with requiring  $N_\mathrm{\mu} \ge 2$ and  leading muon $p_\mathrm{T}>25$ GeV, 
for events with \texttt{triggerIsoMu24==1}. In addition, require that the two muons have opposite charge.
Key figures to show in a presentation:
\begin{itemize}
\item invariant mass of the two muons $m(\mu^{+}, \mu^{-})$ in bins of 0.25 GeV width in the range $[0, 20]$ GeV
\item invariant mass of the two muons $m(\mu^{+}, \mu^{-})$  in bins of 10 GeV width in the   range $[20, 160]$ GeV
\item MET in bins of 10 GeV width
\item jet multiplicity $N_{j}$
\item b-jet multiplicity $N_\mathrm{bj}$
\item event counting statistics summary
\end{itemize}

\subsection{\texorpdfstring{\ttbar\ }{ttbar} cross section in the \texorpdfstring{$\mu e$}{mu e} final state}
The event preselection starts with requiring  $N_\mathrm{\mu} \ge 1$ and  leading muon $p_\mathrm{T}>25$ GeV, at least one electron $N_\mathrm{e} \ge 1$, for events with \texttt{triggerIsoMu24==1} true. 
In addition, require that the two charged leptons have opposite charge. Key figures:
\begin{itemize}
 \item invariant mass of the two leptons $m(l^{+}, l^{-})$  in bins of 1 GeV width in the   range $[0, 160]$ GeV
 \item MET in bins of 10 GeV width
 \item jet multiplicity $N_{j}$
 \item b-jet multiplicity $N_\mathrm{bj}$
 \item event counting statistics summary
\end{itemize}

In this final state we expect significant contribution from the Drell-Yan ($Z/\gamma^{*}$) process, explain why and how the cut on the $m(l^{+}, l^{-})$ might help getting rid of this process.

\subsection{\texorpdfstring{\ttbar\ }{ttbar} cross section in the \texorpdfstring{$\mu + \mathrm{bjet} + \met$}{mu + bjet + MET} final state}
The event preselection starts with requiring  $N_\mathrm{\mu} = 1$ and  leading muon $p_\mathrm{T}>25$ GeV, $N_\mathrm{bj} \ge 1$, for events with \texttt{triggerIsoMu24==1} true. 
Key figures:
\begin{itemize}
\item  muon $p_\mathrm{T}$ in bins of 5 GeV width
\item  MET in bins of 10 GeV width
\item  jet multiplicity $N_{j}$
\item  b-jet multiplicity $N_\mathrm{bj}$
\item  event counting statistics summary 
\end{itemize}

\subsection{\texorpdfstring{\ttbar\ }{ttbar} cross section in the \texorpdfstring{$\mu + 4\mathrm{jet} + 2\mathrm{bjet} + \met$} final state}
The event preselection starts with requiring  $N_\mathrm{\mu} = 1$ and  leading muon $p_\mathrm{T}>25$ GeV, $N_{j} \ge 4$, $N_\mathrm{bj} \ge 2$, for events with \texttt{triggerIsoMu24==1} true. 
Key figures:
\begin{itemize}
\item  muon $p_\mathrm{T}$ in bins of 5 GeV width
\item  MET in bins of 10 GeV width
\item  jet multiplicity $N_{j}$ 
\item  jet multiplicity $N_{j}$ when no cut on $N_{j}$ is placed, but all other cuts are applied
\item  b-jet multiplicity $N_\mathrm{bj}$ 
\item  b-jet multiplicity $N_\mathrm{bj}$ when no cut on $N_{j}$ is placed, but all other cuts are applied
\item  event counting statistics summary
\end{itemize}

\subsection{Reconstruction of the t-quark and W-boson masses}
Reconstruct the $t$ quark and $W$ boson masses in the semi-leptonic $t\bar{t}$ final state, 
requiring  $N_\mathrm{\mu} = 1$ and  leading muon $p_\mathrm{T}>25$ GeV, $N_{j} \ge 4$, $N_\mathrm{bj} \ge 2$, for events with \texttt{triggerIsoMu24==1} true. 
Assume the final state $tt\to WWbb\to \mu\nu qqbb$ as fully resolved, where we have omitted charge and anti-particle notation for simplicity. 
Assume that the first four leading jets can be attributed to $qqbb$. The $qq$ are the two jets that are not b-tagged, while for $bb$ we assign the two jets that pass the b-tagging threshold. 

We interpret the $qq$ pair jets as coming from the hadronic decay of the W boson. Compute the invariant mass distribution of the two $q$ jets $m_\mathrm{qq}$ as well as the invariant mass distribution of the three jet system $m_\mathrm{qqb}$ assuming that is coming from the same parent $t$ quark decay. We don't know which of the two b-jets is the correct one to be paired with the $qq$, i.e., which of the bjets has the same $t$-quark parent as the $q$-jets. Try both combinations and name them $m_\mathrm{qqb_{1}}$ and $m_\mathrm{qqb_{2}}$, where $b_{1}$  and $b_{2}$ is the leading and sub-leading b-jets.
Key figures:

\begin{itemize}
 \item  muon $p_\mathrm{T}$ in bins of 5 GeV width
 \item  MET in bins of 10 GeV width
 \item  jet multiplicity $N_{j}$ 
 \item  b-jet multiplicity $N_\mathrm{bj}$ 
 \item  $m_\mathrm{qq}$ in bins of 10 GeV width
 \item  $m_\mathrm{qqb_{1}}$ in bins of 10 GeV width
 \item  $m_\mathrm{qqb_{2}}$ in bins of 10 GeV width
 \item   $m_\mathrm{qqb_{1}}$ and $m_\mathrm{qqb_{2}}$ stacked in the same histogram (e.g., fill the 2 entries in the same histogram for each event)
 \item  event counting statistics summary
\end{itemize}

\subsection{Charge asymmetry in \texorpdfstring{$pp\to W^{\pm}$}{pp to W+/-}}
Select a $W$ boson control region, find how many of them are $W^{+}$ and $W^{-}$. What do we naively expect from the total charge of the initial state ? What our MC simulation predicts for the charge asymmetry ?

\subsection{Charge asymmetry in \texorpdfstring{\ttbar}{ttbar}}
Select semileptonic \ttbar events, find how many of them are $W^{+}$ and $W^{-}$. What do we naively expect and how it compares with what MC simulation predicts ?

\input{aknow}

\appendix 
\section{Invariant Mass}
To calculate the invariant mass of $X\to A B$ decays, given the four-vectors $p_\mathrm{A}^{\mu} = (E_\mathrm{A}, p_\mathrm{Ax}, p_\mathrm{Ay}, p_\mathrm{Az})$ and
$p_\mathrm{B}^{\mu} = (E_\mathrm{B}, p_\mathrm{Bx}, p_\mathrm{By}, p_\mathrm{Bz})$ we use the square of four-momentum conservation 
$P^{\mu}_\mathrm{X} = p_\mathrm{A}^{\mu} + p_\mathrm{B}^{\mu}$, which gives
\begin{equation}
M^{2}_\mathrm{X}= (E_\mathrm{A} + E_\mathrm{B})^2 - (p_\mathrm{Ax} + p_\mathrm{Ax})^2 - (p_\mathrm{By} + p_\mathrm{By})^2 - (p_\mathrm{Cz} + p_\mathrm{Cz})^2.
\end{equation}
Having computed $M^{2}_\mathrm{X}$ we only need to take its square root to end up to $M_\mathrm{X}$.

\section{Transverse Mass}
To calculate the transverse mass $(m_\mathrm{T})$ in $W\to \ell \nu$ decays, we will work under the assumption that the visible MET is solely due to the transverse momentum of one escaping neutrino. 
We will neglect the muon and the neutrino masses and build their transverse 4 vectors such as they are light-like $(P^2 = 0)$, using only the transverse component of their 
momentum 
\begin{equation}
p^{\mu}= (E, p_\mathrm{x}, p_\mathrm{y}, p_\mathrm{z}) =  (\sqrt{p_\mathrm{x}^2+p_\mathrm{y}^2}, p_\mathrm{x}, p_\mathrm{y}, 0).
\end{equation}
We will sum the two  transverse 4-vectors and calculate the magnitude (mass) of their sum, which is the definition of the $m_\mathrm{T}$. 
Note that is not possible to calculate the ordinary invariant mass of the $m(\mu, \nu)$ system, since  the $p_\mathrm{z}$ of the $\nu$ is unknown. 
The $m_\mathrm{T}$ is the closest quantity we could built to the invariant mass of the system of two particles, having as endpoint the $m(\mu, \nu)$ 
and being itself also invariant. See also Sec.~49.6 of \href{https://pdg.lbl.gov/2022/reviews/rpp2022-rev-kinematics.pdf}{PDG2022}. 

\section{Error bars in histogram bins}
Histograms are the most usual way to quickly estimate the shape of the underlying probability density function of an observable. 
Suppose that the random variable $X$ we wish to measure takes continuous values, as for example the $\pt$ of a muon.
We count how many events have a muon within a certain $\pt$ range (bin of our histogram) and populate the event content of that particular bin. 

We are interested to learn about the probability $p_{j}$ that $X$ is observed inside the boundaries  
$x_\mathrm{min,j} < x <x_\mathrm{max,j}$ of the ${j}$-th bin. 
In the limit of $N\to \infty$, an estimator of the desired probability is the $\hat{p_{j}} = n_{j}/N \approx p_{j}$, 
where N is the total number of events we have and $n_{j}$ the subset of those that are observed within the bin range $[x_\mathrm{min,j}, x_\mathrm{max,j})$\footnote{
By convention, bin intervals are usually {closed ``[''} on the lower end and {open ``)''} on the higher.}.

We may regard each bin of a histogram as an independent experiment, governed by the binomial probability 
\begin{equation}
\label{eq:binomial}
P(n_{j}) = \frac{N!}{n_{j}! (N-n_{j})!} p_{j}^{n_{j}} (1-p_{j})^{N-n_{j}}
\end{equation}
as the event will either belong in bin ${j}$ or not. Like when we flip a coin it is either heads or tails. But there, the two outcomes are equiprobable while for our observable
one of the two outcomes might be very rare.
The variance of $n_{j}$ is 
\begin{equation}
\sigma_{n_{j}}^{2} = E(n_{j}^2) - E(n_{j})^2 = N p_{j} (1-p_{j}). 
\end{equation}
When our random variable $X\sim p(x)$ is {\em continuous} and distributes according to the $p(x)$ probability density function, we have
\begin{equation}
E(n_{j}) = N \int_{x_{min,j}}^{x_{max, j}} p(x) dx = N p_{j}
\end{equation}
with $E(n_{j})$ denoting the theoretically expected value of $n_{j}$.
In the limit of $N\to \infty$ we can always chose the bin boundaries with fine segmentation (fine binning), such as $p_{j}\to 0$ while at the same time $N p_{j}$ remains 
finite and equal to $n_{j}$.
In this limit, Eq.~\ref{eq:binomial} \href{https://en.wikipedia.org/wiki/Poisson_limit_theorem}{is reduced to a Poisson},
\begin{equation}
P(n_{j}) = \frac{\mu_{j}^{n_{j}}}{n_{j}!}\exp^{-\mu_{j}}
\end{equation}
with $\mu_{j} = N p_{j}$ and $\sigma_{n_{j}}^{2} = N p_{j} (1-p_{j}) = N p_{j} =  \mu_{j}$, since $1-p_{j} \approx 1$ when $p_{j} \to 0$.
Thus, the bin entries distribute with Poisson probability mass function having $\sigma_{n_{j}}^{2} = \mu_{j}$. 
Unfortunately, the Poisson confidence intervals are not simple in calculation and one has to use the quantile of the chi-square distribution (in \texttt{ROOT} it's the \texttt{TMath::ChisquareQuantile}) 
which give asymmetric error bars. 
\begin{verbatim}
float statErrorN(float x){return x - 0.5*TMath::ChisquareQuantile(0.3173/2,2*x);}
float statErrorP(float x){return 0.5*TMath::ChisquareQuantile(1-0.3173/2,2*(x+1))-x;}
\end{verbatim}
Doing error propagation with asymmetric error bars is tedious. People sometimes take the largest of the up/down error bar and symmetrize it calling this procedure conservative.

We can further simplify life making the binning such as the $E(n_{j})>20$, while at the same time  $p_{j}\to 0$ and $N\to\infty$.
We are one step before saying that $\delta{n_{j}} = \sigma_{n_{j}} = \sqrt{E(n_{j})} \approx \sqrt{n_{j}}$ and treat it as a {\em Normal} symmetric confidence interval of 68\% confidence level since for $E(n_{j})\gg20$ the Poisson distribution \href{http://www.stat.ucla.edu/~dinov/courses_students.dir/Applets.dir/NormalApprox2PoissonApplet.html}{is well approximated by the Normal distribution}. 
For {\em continuous} random variables and very large number of events, we can always chose fine binning such as $p_{j} \to 0$, thus rendering the above considerations reasonable.
However, an extra leap is needed to use the {\em observed number of events} $n_{j}$ as an estimator of the {\em unknown} $E(n_{j}) = \mu_{j} = N p_{j}$.

In the realm of statistics \href{https://www.science20.com/quantum_diaries_survivor/those_deceiving_error_bars-85735}{it can be disputable} to attach error bars on the 
observed data of a counting experiment. Unarguably, we are $100\%$ sure how many we have already observed.
My personal take on the subject. For displaying error bars in a plot and back to the envelope calculations, using $\sqrt{n}$ as the error bar of $n$ {\em observed} events 
is not unreasonable~\footnote{In the Gaussian limit large-N, the confidence belt is symmetric, i.e in the Neyman construction, the width of the confidence belt as obtained from the observed value is the same as that of the pdf of the true value.
}, provided that our bins have at least 20 events as content and we do not speak about $0\pm0$ or $1\pm1$. 
The number of events we have is related to the statistical precision of our data and the $\sqrt{n}$ prescription is in many times helpful when visualizing the data. 
However, we should keep in mind 
the error bars displayed in figures should not necessarily, and as a matter of fact are very often not, propagated as such to the final result.

\section{Poisson processes}
The number of phone calls $(n)$ a help line receives over a fixed time interval (e.g., per 5 minutes) is following the Poisson distribution, 
\begin{equation}
\label{eq:poisson}
P(n) = \frac{\mu^n}{n!}\exp^{-\mu}
\end{equation}
with $\mu$ being the mean number of phone calls received per time interval.
It is easy to show that the maximum likelihood estimator of $\mu$ is just the arithmetic mean (average) rate of phone calls per time interval.
Same goes for the number of nuclear decays recorded in a certain time interval by a \texttt{Geiger-M\"uller} detector, 
or for the number of events recorded by the CMS detector satisfying specific selection criteria for a fixed amount of integrated luminosity. 

You may wonder what is common between a telephone center and nuclear and particle physics. The answer is,
\begin{itemize}
\item we have {\em many} clients ($pp$ collisions) each of them having the {\em very same small probability} $p \to 0$ calling the help center (passing the event selection criteria).
\item each client ($pp$ collision) is {\em independent} from the others and {\em memory-less}, i.e., future $pp$ collisions don't care what was the outcome of past $pp$ collisions and the probability that a client calls the help center is {\em constant} and {\em the same} for all clients.
\end{itemize}
The above conditions bring us to close to the \href{https://en.wikipedia.org/wiki/Poisson_limit_theorem}{Poisson limit} of Eq.~\ref{eq:binomial}. 
Verify numerically the that Eq.~\ref{eq:binomial} is well approximated by Eq.~\ref{eq:poisson} when $N\to\infty$ and $p\to 0$ with $Np = \mu$ giving your own values to $p$ and $N$.

%% file: aknow.tex
\section{Acknowledgments}
I would like to thank Isabell Melzer-Pellman for organizing and inviting me to the Introduction to the Terascale School in DESY 2023 and to Andreas Meyer for reading and providing precision comments to the present document.